\documentclass[acmsmall,screen,nonacm]{acmart}
\usepackage{array}
\usepackage{tcolorbox}
\usepackage{placeins}
\newcommand{\leadparagraph}[1]{\noindentparagraph{\textnormal{\textbf{#1}}}}
\setcopyright{none}
\begin{document}
\title[From Software History to Executable Regression Tasks]{From Software History to Executable Regression Tasks: A System for Evaluating Software Agents}

\author{Ruikai Shi}
\authornote{These authors contributed equally to this work.}
\affiliation{\institution{Peking University}\city{Beijing}\country{China}}
\email{rkshi25@stu.pku.edu.cn}

\author{Zhengran Zeng}
\authornotemark[1]
\affiliation{\institution{Peking University}\city{Beijing}\country{China}}
\email{zhengranzeng@stu.pku.edu.cn}

\author{Rui Xie}
\authornote{Corresponding authors: Rui Xie, Shikun Zhang, and Wei Ye.}
\affiliation{\institution{Peking University}\city{Beijing}\country{China}}
\email{ruixie@pku.edu.cn}

\author{Shikun Zhang}
\authornotemark[2]
\affiliation{\institution{Peking University}\city{Beijing}\country{China}}
\email{zhangsk@pku.edu.cn}

\author{Wei Ye}
\authornotemark[2]
\affiliation{\institution{Peking University}\city{Beijing}\country{China}}
\email{wye@pku.edu.cn}

\renewcommand{\shortauthors}{Shi et al.}
\authorsaddresses{}
\thanks{Authors' email addresses: Ruikai Shi (\href{mailto:rkshi25@stu.pku.edu.cn}{rkshi25@stu.pku.edu.cn}); Zhengran Zeng (\href{mailto:zhengranzeng@stu.pku.edu.cn}{zhengranzeng@stu.pku.edu.cn}); Rui Xie (\href{mailto:ruixie@pku.edu.cn}{ruixie@pku.edu.cn}); Shikun Zhang (\href{mailto:zhangsk@pku.edu.cn}{zhangsk@pku.edu.cn}); Wei Ye (\href{mailto:wye@pku.edu.cn}{wye@pku.edu.cn}).}

\begin{abstract}
As software agents increasingly generate and modify code, evaluating their
ability to review changes and discover defects becomes important. Such evaluation
needs executable project environments for investigation and reliable defect
references supported by reproducible behavior. Existing code review benchmarks
provide limited support for this combination. We present a system
that constructs executable regression instances from historical introducing--fixing
commit pairs and developer tests. The system extracts candidate tests from
fixing-side changes, then uses constrained agents to transplant them and configure
version-specific environments while preserving production code and test
expectations. Independent replay and evidence-based review verify whether tests
that fail on the introducing version and pass on the fixing version support the
proposed regression relationship. Accepted instances retain both versions'
replay recipes, validated tests, and execution evidence. Repair-supported analysis
and human review add root-cause locations, tested repairs, and causal explanations.
These instances support bug detection and localization in code review, and
further enable bug repair and regression-test generation on the same defects.
The system constructs 172 validated instances from 522 candidates across
11 Python projects. We evaluate three software agents with ten models and
additional reasoning-effort settings on 83 manually selected, enriched instances
across the three tasks. Precise root-line localization remains challenging,
while baseline combinations achieve repair and regression-test generation
success rates of up to 68.7\% and 77.1\%, respectively. Leading configurations
differ by task, and higher task performance does not necessarily yield greater
API-cost efficiency. Our approach provides reusable executable evidence for
studying agents' defect discovery, repair, and testing capabilities.
\end{abstract}

\maketitle
\pagestyle{plain}
\thispagestyle{plain}
\section{Introduction}
\label{sec:introduction}

As LLM-based agents take on more code generation and modification
tasks~\cite{yang2024sweagent}, evaluating and improving their code review and
defect discovery capabilities becomes increasingly important. Finding defects
in a change can require executing the program, constructing inputs, and checking
suspected behavior. Evaluation therefore needs executable project environments
that support investigation and reliable reference answers identifying the
defects and their locations.

SWR-Bench, CodeFuse-CR-Bench (SWE-CARE), and AACR-Bench provide pull requests,
repository context, and review
annotations~\cite{zeng2026swrbench,guo2025codefusecr,zhang2026aacrbench}.
Their published construction workflows do not describe systematic reconstruction
and validation of project execution conditions for each reviewed revision.
Repository snapshots and general Agent containers still require historical
dependencies and test support to run the affected behavior reliably.
Reliable evaluation also needs reproducible behavioral evidence supporting the
reference defect. We address these needs by constructing executable regression
instances from software history.

Software history records introducing changes, repairs, and developer tests.
Constructing executable instances presents three challenges.
Fixing-side test changes may pass on both
versions or exercise unrelated behavior. Historical versions may require different
dependencies and test support that must be reproducible beyond an Agent session.
Fail-to-pass outcomes must reflect proposed code changes rather than incompatible test support or unintended validation edits.

We present \textbf{EReg-Factory}, a system for constructing executable regression
data from candidate introducing--fixing pairs. The system extracts fixing-side
candidate tests for constrained Validation Agents to transplant and execute in
version-specific environments. Operation Judges check Agent actions; an environment
replayer independently reruns tests; and a Regression Verifier connects paired
outcomes to code changes. Instances retain two replay recipes, validated tests,
execution evidence, and explanations. Repair-supported analysis and human review
add root locations, tested repairs, and causal explanations. These instances
support bug detection and localization in code review, where Agents examine
the introducing change without access to later fixes or target regression tests.
Retained repairs and behavioral evidence further support bug repair and
regression-test generation. These three related tasks are evaluated independently
on the same defects.

Across 522 candidates from 11 Python projects, EReg-Factory yields 172 validated
regression instances. We evaluate three Agents with ten models and additional
reasoning settings on 83 manually selected enriched instances across the three
tasks. Precise root-cause localization remains challenging, while repair and
regression-test generation show promise. Leading models differ by task, and
higher performance does not necessarily provide
greater benefit per dollar.

\begin{samepage}
Our contributions are:
\begin{itemize}
  \item \textbf{A regression construction system} combining paired environment
  construction, independent replay, and evidence review to validate historical
  introducing--fixing relationships and retain reproducible execution conditions.
  \item \textbf{Instance enrichment for three tasks} using reviewed root locations,
  repairs, and explanations to support code review through bug detection and
  localization, alongside repair and regression-test generation.
  \item \textbf{An empirical study} characterizing construction outcomes and
  comparing Agents, models, and reasoning configurations by task performance
  and cost efficiency.
\end{itemize}
\end{samepage}

\section{Motivation and Instance Definition}
\label{sec:motivation}

A Checkmk regression illustrates why a related commit pair alone is insufficient
to construct executable regression data.

\begin{figure}[htbp]
\centering
\includegraphics[width=0.90\linewidth]{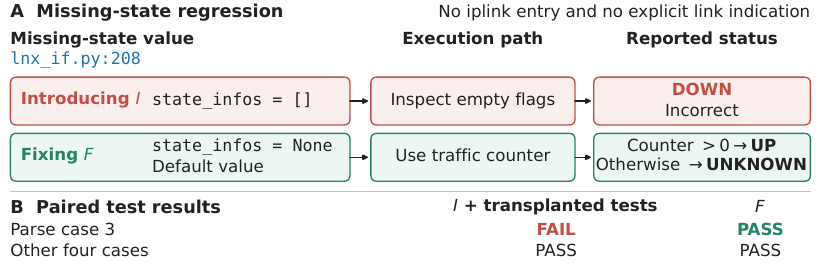}
\caption{Checkmk's missing-state regression: (A) abbreviated code and
(B) paired test results.}
\Description{The introducing version supplies an empty state list at line 208,
causing a DOWN result without explicit link information. None allows the
existing traffic-counter fallback to return UP or UNKNOWN. One parameter case
fails only in the introducing version; four others pass in both versions.}
\label{fig:checkmk-mechanism}
\end{figure}

\leadparagraph{A missing-state regression.}
When neither an iplink record nor an explicit link indication is available,
Checkmk uses traffic counters to report UP or UNKNOWN. A refactoring replaced
\texttt{None} with an empty list for missing state. Because the downstream check
treats any non-\texttt{None} sequence as available state information, this change
produces DOWN and bypasses the fallback. The fix restores \texttt{None} while
retaining the refactoring (Figure~\ref{fig:checkmk-mechanism}).

After transplanting the fixing-side candidate tests, independent execution yields
four passes and one failure on the introducing version, versus five passes on
the fixing version. Only the fail-to-pass case is a regression witness.

\leadparagraph{From historical changes to executable evidence.}
The example motivates three requirements: selecting tests that expose the
affected behavior, reconstructing each version's execution environment, and
linking the observed difference to the introducing and fixing changes rather
than incompatible test support.

\leadparagraph{Executable regression instances.}
Let $I$ and $F$ denote candidate introducing and fixing commits. We seek a
nonempty set of regression cases $T^*$ from tests changed in $F$, such that
\begin{equation}
  O_I(t)=\mathrm{FAIL}\quad\text{and}\quad O_F(t)=\mathrm{PASS},
  \qquad \forall t\in T^*.
\end{equation}
We measure $O_I$ with transplanted tests and necessary non-production adaptations while preserving production code at $I$, and $O_F$ at $F$, each in its own environment. Code and execution evidence must link these outcomes to the
target functional or compatibility regression. Each instance stores a commit pair, two replay recipes, validated tests, execution evidence and behavioral explanations.

Section~\ref{sec:construction} constructs these instances;
Section~\ref{sec:enrichment} enriches them with root annotations and tested repairs
for the three evaluation tasks.

\section{EReg-Factory: Regression Instance Construction}
\label{sec:construction}

\begin{figure}[htbp]
\centering
\includegraphics[width=\linewidth]{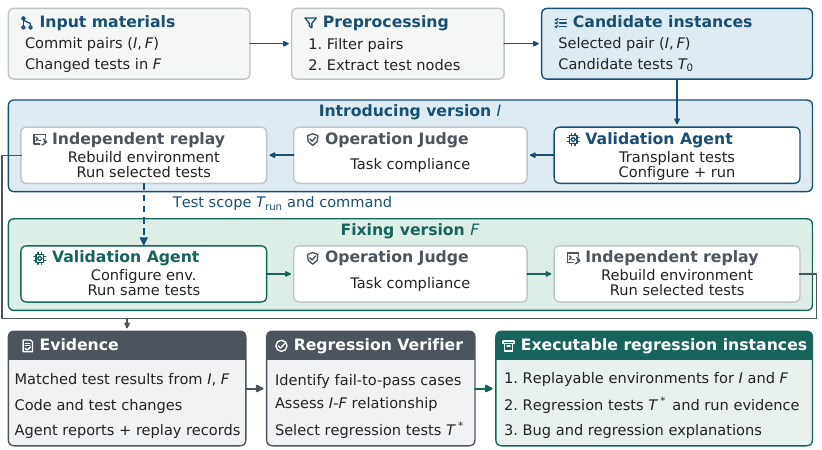}
\caption{The EReg-Factory construction pipeline. The dashed arrow
transfers test scope and command; non-accepted branches are omitted.}
\Description{The top row proceeds from Input materials, grouping commit pairs and
changed tests in their corresponding fixing commit F, through Preprocessing
with pair filtering and test-node extraction, to Candidate instances containing a selected
pair I,F and candidate tests T0. A straight downward arrow feeds the introducing-side
Validation Agent. This lane runs right to left from Agent to operation Judge to
independent replay; the fixing-side lane runs left to right. A separate straight
downward dashed arrow transfers only test scope and command to the fixing-side
Agent. The introducing-side Agent
transplants the selected test files and configures its environment;
the fixing-side Agent configures a separate environment without source or test
edits. Each attempt passes an operation Judge before independent replay in a
fresh container. The introducing side supplies the fixing side's final test
scope and command, not its environment or results. The Regression Verifier
receives a separate evidence collection containing matched test results, code and
test changes, Agent reports and replay records. It identifies fail-to-pass cases,
assesses the introducing--fixing relationship, and selects supported regression
tests. Accepted instances retain two replayable environments, execution
evidence, supported regression tests T star, and bug and regression explanations.
Root-cause enrichment and evaluation task design are subsequent stages, not shown.}
\label{fig:system-overview}
\end{figure}

EReg-Factory constructs executable regression instances from candidate
introducing--fixing commit pairs through paired validation, independent
environment replay, and regression verification. This section describes that
core construction process.
Figure~\ref{fig:system-overview} summarizes the process from candidate acquisition
to a validated, independently replayable regression instance.

\subsection{Candidate Relationships and Test Scope}
\label{sec:acquisition}

\begin{figure}[htbp]
\centering
\includegraphics[width=\linewidth]{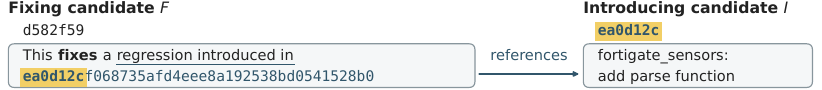}
\caption{Identifying a candidate commit pair from a commit message.}
\Description{Commit d582f59 states that it fixes a regression introduced in
ea0d12cf068735afd4eee8a192538bd0541528b0. An arrow follows this explicit SHA
reference to introducing candidate ea0d12c.}
\label{fig:message-link}
\end{figure}

Candidate acquisition is replaceable: other mining methods or curated sources
can supply pairs, provided that both commits and fixing-side test changes
are available. We use GH Archive, which records public GitHub activity in
hourly archives~\cite{gharchive}. Its cross-repository history enables mining
without selecting repositories individually in advance. We mine its 2023--2025 PushEvents across
repositories. Case-insensitive matching identifies repair
(\texttt{fixes}, \texttt{correct}, \texttt{resolve}) or causal terms
(\texttt{regression}, \texttt{introduced by}, \texttt{broken since}), followed
by a full 40-character hexadecimal SHA. The message links fixing candidate $F$
to introducing candidate $I$ through an explicit SHA reference, without
cross-message similarity matching
(Figure~\ref{fig:message-link}). Filters remove dependency updates,
routine merges, CI, documentation, formatting, releases and backports;
repository and test-change screening retains Python projects with pytest
indications and candidate test files.
Following this initial rule-based screening, all candidate relationships undergo
rigorous manual secondary screening to confirm their credibility and provide
reliable inputs for subsequent validation.

To further screen candidate inputs, we examine the pull requests associated
with each fixing candidate. Fixing candidates pass when any associated pull request is merged; those
without requests also proceed to later validation, accommodating direct commits.
Those with only unmerged requests are excluded. For eligible fixes referencing one introducing commit, we select the latest event: later attempts may complete an earlier repair. Paired execution tests this relationship.

We treat tests changed by a fixing commit as candidate regression tests because
they may express the behavior targeted by the fix and its expected outcome.
Pytest tests check program behavior through given inputs and assertions on
expected results, providing a basis for reusing the same behavioral check across
versions. In subsequent stages, we preserve these inputs and assertions and
transplant the complete fixing-side tests to the introducing version, checking
the earlier version against the same expectations after necessary adaptation. Changes confined to
assertions or expected results can therefore also yield candidates.

We map changed locations in test code onto Python abstract syntax trees to
identify the corresponding test functions or methods and generate pytest nodes,
forming candidate scope $T_0$. If no eligible nodes are identified, we discard
the candidate commit pair. For retained candidates, later stages configure the environment and adapt test support to verify executability and behavioral differences between versions, establishing valid regression tests.

\subsection{Validation on the Introducing Version}
\label{sec:paired}

Introducing-side validation tests whether fixing-side candidates execute on $I$
and expose a failure. Historical dependencies and test infrastructure can cause
failures unrelated to a regression. A \textbf{Validation Agent} prepares and runs
the tests, an \textbf{operation Judge} checks compliance, and independent replay
provides inspectable, repeatable execution evidence.

\begin{figure}[htbp]
\centering
\includegraphics{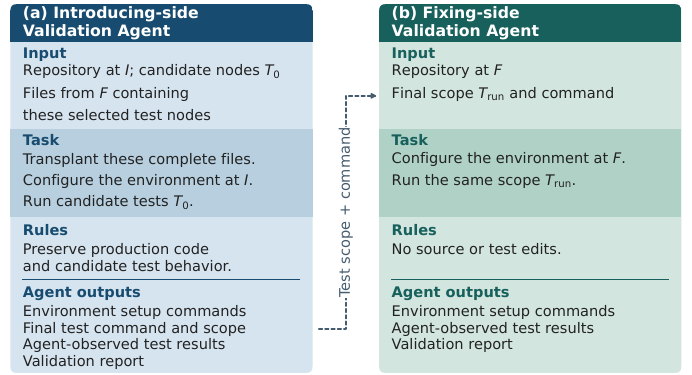}
\caption{Introducing- and fixing-side Validation Agent tasks; the dashed arrow transfers final test scope and command.}
\Description{Two aligned panels summarize inputs, tasks and outputs. The
introducing-side Validation Agent receives the repository at I, selected candidate nodes
T0, and the files from F containing those nodes. It transplants these complete
files, configures an environment and runs the candidates, preserving production
code and candidate test behavior. It reports the final scope and command. A
dashed arrow carries that information to the fixing-side Validation Agent, which configures
its own environment at F and runs the same scope Trun without source or test
edits. Both sides retain their own environment setup commands, Agent-observed
test results and validation report. The introducing side additionally outputs
the final test command and scope. These observations precede independent replay
and final review. The arrow transfers only test scope and command, not results
or environment setup commands. Their environments are independent.}
\label{fig:validation-agents}
\end{figure}

We build the Validation Agent on SWE-agent~\cite{yang2024sweagent}, adding task
instructions and tools for repository inspection, dependency configuration,
candidate test transplantation and execution, and evidence collection. It
observes test behavior rather than repairing the program and produces a
structured report. Figure~\ref{fig:validation-agents} summarizes both sides' tasks
and instructions.

The introducing-side Validation Agent receives the repository, the introducing and fixing
commit identifiers, candidate test paths and the specified pytest nodes $T_0$.
It applies to $I$ the complete fixing-side test files containing the nodes
extracted in Section~\ref{sec:acquisition}, configures the required environment,
and executes those candidate nodes $T_0$. Existing non-production fixtures, helpers, test
data and test configuration may be restored or adapted to support execution,
but candidate inputs, assertions, expected results, parameters and skip behavior
must remain unchanged. Production implementation cannot be copied or edited,
even to resolve an import, because doing so could import the fix under
investigation. If some candidates cannot execute within these constraints,
the Agent may narrow the scope to $T_{\mathrm{run}}\subseteq T_0$ and report
the omitted tests. Its final report records the environment preparation,
self-contained test command, exit code and observed outcome.

\begin{figure}[htbp]
\centering
\includegraphics[width=\linewidth]{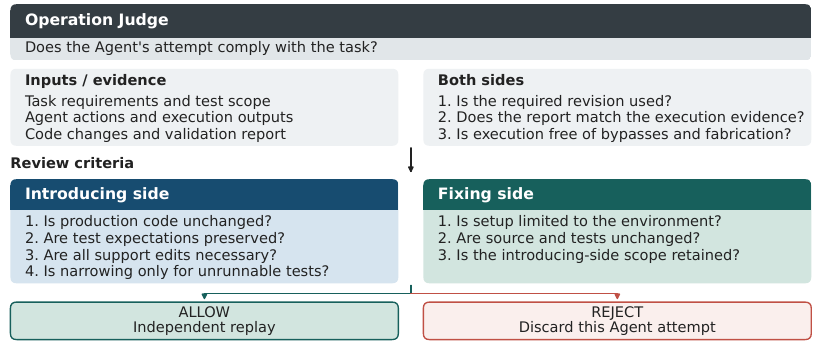}
\caption{Operation compliance checking before independent replay.}
\Description{The Judge receives task requirements and test scope, Agent actions
and execution outputs, code changes and the validation report, but no subsequent
independent replay results. Introducing-side criteria preserve production
implementation and candidate test expectations, allow necessary nonproduction
support adaptations, and permit narrowing only for tests that cannot run.
Fixing-side criteria require environment-only preparation without source or test
edits and execution of the introducing-side test scope. Both sides require the
specified revision, faithful reporting, and no test bypass or fabricated
behavior. ALLOW leads to independent replay; REJECT discards that Agent attempt.}
\label{fig:validation-judge}
\end{figure}

\begingroup\widowpenalties 3 10000 10000 0
Before independent replay, an LLM-based Judge reviews task compliance using
the task requirements, Agent execution record, code changes and validation
report (Figure~\ref{fig:validation-judge}). It allows supporting adaptations
needed to run the tests while requiring the candidate tests' behavioral
expectations and the production implementation to remain unchanged.
Runs that pass this review proceed to independent replay.
We use GPT-5.6 Sol with reasoning effort set to \texttt{max} for the Judges
on both the introducing and fixing sides.
\par\endgroup

\leadparagraph{Independent replay.}
\label{sec:replay}
Environment construction is exploratory, involving dependency installation,
temporary inputs, environment variables and repository build commands; neither
a package list nor the final test command captures all these conditions. Our
\textbf{environment replayer} converts the recorded configuration process into a
recipe executable in a fresh container, avoiding reliance solely on the Agent's
account or temporary session. SetUpAgent and Repo2Run reuse Agent exploration
for setup~\cite{vergopoulos2025setupagent,hu2025repo2run}; our replayer couples this
process with the historical source state and candidate test scope.

Replay starts from the target commit and extracts environment configuration
operations and their required temporary inputs from the execution trace,
preserving their order and necessary environment settings. Environment and
source state are reconstructed separately: setup and build commands restore
dependencies and generated artifacts, while a patch applies test transplantation
and permitted supporting changes, avoiding duplicate application of generated
artifacts. The replayer then reconciles the final report with the execution
record to determine the test command and runtime settings, so that replay
corresponds to the Agent's final reported experiment rather than an exploratory
attempt.

We also retain dependency information and recorded versions for Python
packages, system packages and build tools used to construct the Agent's
environment. When replay exposes missing dependencies during
test startup, collection or setup, the replayer uses these records to install
the corresponding package versions and reruns the same test command, rather
than modifying the program under test to eliminate failures. Replay preserves
per-case results and failure output for behavioral comparison, regression
verification and re-examination.

\subsection{Validation on the Fixing Version}
\label{sec:fixing-validation}

Fixing-side validation checks that the tests exercised on $I$ pass on $F$,
resolving the observed failures. The Validation Agent receives the repository
at $F$ and the introducing side's final test command and exercised scope,
including any narrowed subset (Figure~\ref{fig:validation-agents}). It configures
its own environment and runs the existing tests without changing source code
or tests, recording the environment, command and outcomes in a structured report.

The fixing-side Judge reviews the execution evidence under this
environment-only constraint, checking that preparation preserves the program,
tests and required test scope. Runs that pass proceed to independent
validation with the environment replayer described in Section~\ref{sec:replay}.
Replay uses the fixing-side Validation Agent's own configuration and final command in a fresh
container. The two sides therefore share a test scope, while allowing the
dependency versions and runtime settings needed by each historical version
to differ.

Paired replays yield independently observed outcomes for the same tests. A
fail-to-pass difference alone does not establish the proposed introducing--fixing
relationship; Section~\ref{sec:regression-review} examines both executions, failure
causes and commit changes.

\subsection{Reviewing the Regression Relationship}
\label{sec:regression-review}

We use an LLM-based \textbf{Regression Verifier} to combine evidence from both validation
stages and assess whether candidate tests support the proposed
introducing--fixing relationship. Our implementation uses GPT-5.6 Sol with
reasoning effort set to \texttt{max}. We align the independently replayed outcomes by individual test
case and examine those that fail on $I$ and pass on $F$. We assess individual parameterized cases, rather than entire test functions, as regression evidence.

\begin{figure}[!htbp]
\centering
\includegraphics[width=\linewidth]{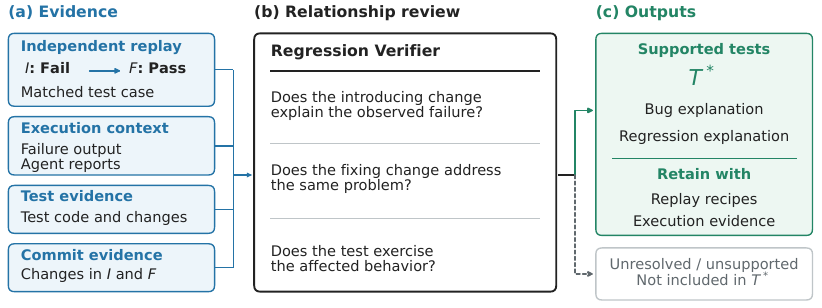}
\caption{Evidence-based regression verification.}
\Description{A left-to-right diagram. Independent replay of a matched test case
fails at I and passes at F. Failure output, Agent reports, test code and changes,
and both commit differences feed the Regression Verifier. Three questions ask whether
the introducing change explains the failure, the fixing change addresses the
same problem, and the test exercises the affected behavior. Outputs include
supported tests, a bug explanation covering its introduction and repair, and
a regression explanation connecting the tests to that behavior. These are
retained with replay recipes and execution evidence. A secondary branch
excludes unresolved or unsupported cases from T*.}
\label{fig:validation-reviewer}
\end{figure}

The verifier receives both Agent reports, paired replay results and failure
output, relevant test code and its changes, and the introducing and fixing
commit differences (Figure~\ref{fig:validation-reviewer}). It examines whether
the change in $I$ explains the observed failure, whether $F$ addresses the same
problem, and whether the test exercises the affected behavior. Unlike the operation Judge's compliance check, this assessment links code changes, test expectations and paired outcomes to distinguish supported regressions from unrelated behavior, environment failures or incompatible test transplantation.

The verifier produces a structured assessment comprising supported regression
tests $T^*$, a bug explanation describing the defect and how it was introduced
and repaired, and a regression explanation describing the behavior checked by
the tests and why their outcomes support that relationship. Unresolved or
unsupported cases are excluded from $T^*$. We retain the supported tests and
explanations with both replay recipes and execution results. These instances
are then enriched with root-cause annotations and task-specific evidence for
evaluation.

\section{Instance Enrichment and Evaluation Task Design}
\label{sec:enrichment}

EReg-Factory constructs executable regression instances with validated tests
and environments for both versions (Section~\ref{sec:construction}). We enrich these
instances with root-cause annotations and tested repairs, then use the resulting
evidence to design three complementary evaluation tasks: bug detection and
localization, bug repair, and regression-test generation.

\subsection{Repair-Supported Root-Cause Annotation}
\label{sec:root-annotation}

\begin{figure}[htbp]
\centering
\includegraphics[width=\linewidth]{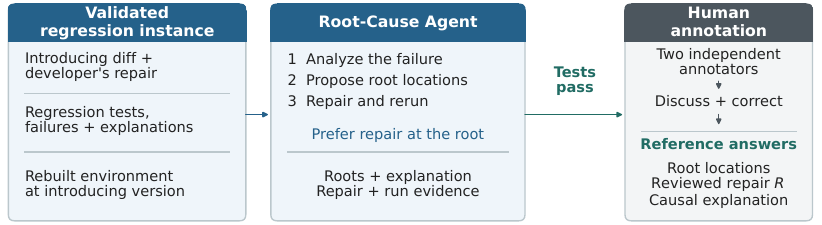}
\caption{Repair-supported root-cause annotation.}
\Description{Three stages flow left to right. A validated instance supplies
introducing and fixing changes, regression tests, failures and explanations,
and a rebuilt introducing environment. The Root-Cause Agent analyzes the
failure, proposes root locations, and repairs and reruns tests, with root-site
repair encouraged. A tests-pass condition gates the handoff of root proposals,
explanations, repairs and execution evidence to two independent annotators.
They discuss and correct the annotations and review the repair, retaining
root locations, a reviewed repair and a causal explanation for evaluation.}
\label{fig:root-annotation}
\end{figure}

Validated regression tests expose affected behavior but do not directly locate
its root cause. Repair locations need not coincide with root-cause locations:
a repair may adjust downstream handling or add a guard to eliminate a failure
whose cause lies elsewhere. We combine repair-based analysis with human annotation to locate the regression's causes in the introducing change, providing reference locations and explanations for evaluation (Figure~\ref{fig:root-annotation}).

The environment replayer reconstructs the introducing version with the
accepted regression tests and confirms their failure. Our
\textbf{Root-Cause Agent} is based on SWE-agent~\cite{yang2024sweagent} and uses DeepSeek
V4 Flash. It receives the prepared environment,
read-only access to the introducing and fixing changes, and the preceding
stage's validated regression tests, failure output, bug explanation and
regression explanation. Together, these provide the context for analyzing
how the problem arose, why the tests expose it, and how the fix changes the
affected behavior. The Agent is instructed to first analyze
the failure and propose root locations, then implement its own repair.
It is encouraged to repair at the identified root cause rather than mechanically
reproduce the developer's repair. It preserves the tests and their execution
conditions and reruns the tests in the same container.
When the tests pass, we retain the proposed source-line root locations, repair
patch $R$ and causal explanation for subsequent manual annotation and review.
The three reviewed artifacts jointly support subsequent evaluation.

Two master's students in computer science \textbf{independently assess all three
artifacts}: whether the root locations are accurate, whether $R$ addresses the
stated problem, and whether the causal explanation agrees with the source and
execution evidence. They annotate and correct locations and explanations,
review and confirm the repair, and resolve disagreements through discussion
to establish the final reference answers. Root locations support localization
comparison and test-coverage evaluation; $R$ provides the repaired state for
regression-test generation; and causal explanations inform task descriptions
and behavior specifications.

\subsection{Evaluation Task Design}
\label{sec:task-construction}

We turn validated and enriched regression instances into three complementary
tasks: finding a problem in a code change, repairing its behavior, and writing
a test that exposes it (Table~\ref{tab:task-protocol}). The tasks share a
behavioral basis but are evaluated independently. Each provides task-specific
inputs to the Agent while retaining reference answers with the evaluator.

\begin{table}[htbp]
\caption{Inputs, outputs and evaluation criteria for the three tasks.}
\label{tab:task-protocol}
\begingroup
\small
\setlength{\tabcolsep}{3.5pt}
\renewcommand{\arraystretch}{1.0}
\begin{tabular}{@{}>{\raggedright\arraybackslash}p{0.12\linewidth} *{3}{>{\raggedright\arraybackslash}p{\dimexpr(.88\linewidth-21pt)/3\relax}}@{}}
\toprule
 & \textbf{Detection / localization} & \textbf{Repair} & \textbf{Test generation} \\
\midrule
Input & Parent and $I$; change description and diff.
& Repository at $I$; symptom description.
& Parent and $I$; diff, behavior specification and test support. \\
\addlinespace[2pt]
Output & Bug judgment; source lines.
& Repair patch.
& Regression tests; support code. \\
\addlinespace[2pt]
Evaluation & Match reported and root-cause lines.
& Test the Agent-repaired version.
& Same tests on $I$ and $I+R$. \\
\addlinespace[2pt]
Criteria & Bug-judgment correctness; line-level agreement.
& All target failures resolved; baseline passes preserved.
& Behavioral fail-to-pass; separate root-line coverage. \\
\bottomrule
\end{tabular}\par
\smallskip
{\footnotesize\raggedright
Introducing version: $I$; validated repair: $R$. Deleted roots use parent lines.
Checks use the evaluated scope. Import/collection errors are excluded;
root coverage is not a pass condition.\par}
\endgroup

\end{table}

\leadparagraph{Bug detection and localization.}
This task assesses whether an Agent can identify a potential problem when
examining a code change. We provide the introducing version, its parent,
the change description and diff, allowing analysis of both added and deleted behavior without
later fixes or regression-test evidence. The Agent judges whether a bug is
present and identifies the responsible source lines without changing the
repository. We compare the reported locations with the instance's root-cause lines.

\leadparagraph{Bug repair.}
Following an evaluation pattern similar to SWE-bench~\cite{jimenez2024swebench},
this task assesses whether an Agent can repair a problem from a symptom
description. The Agent receives this description and the prepared introducing
repository, analyzes the relevant code, and produces a repair without access to the target regression
tests. We then run the relevant tests to determine whether the repair resolves
the targeted failures while preserving previously passing cases within the
evaluated scope. If any targeted failure remains or a previously passing case
fails, the repair is unsuccessful.

\leadparagraph{Regression-test generation.}
This task assesses whether an Agent can turn a behavioral requirement into
a test that detects the regression. The Agent receives a behavior specification,
the parent and introducing versions, and their diff. Ordinary tests and reusable
support remain available. The Agent analyzes the specified behavior and designs
its own test inputs, assertions and implementation without changing production
code. Its tests must
fail on the introducing version and pass on that version with the validated
repair $R$ applied. The failure must expose a behavioral difference, rather
than an import or test-collection error. The generated tests should also cover
as many of the regression's root-cause lines as possible.

To assess repair and test generation more independently of an Agent's ability
to discover and localize a problem from scratch, we provide concise descriptions
of the target behavior. We use GPT-5.6 Sol with reasoning effort set to
\texttt{max} to generate repair symptom descriptions and test behavior
specifications from the instances' validated regression tests and execution
evidence; the latter also use the introducing change. We manually review these
descriptions for fidelity to the evidence and answer leakage, retaining necessary
trigger conditions and behavioral requirements without revealing root locations,
repair solutions or reference test implementations. The Agent must still analyze
the repository to complete each task.

\section{Study Design}
\label{sec:study}

We study the regression data constructed by EReg-Factory and Agent performance
on the three tasks defined in Section~\ref{sec:task-construction}.

\subsection{Research Questions}
\label{sec:research-questions}

\leadparagraph{RQ1: What executable regression data can EReg-Factory construct?}
We examine construction output, project distribution, and the root-cause and
test evidence supporting the three tasks.

\leadparagraph{RQ2: How do Agents and models differ across the three tasks?}
We compare Agents and models on the same instances to examine whether their
relative strengths carry across localization, repair and regression-test generation.

\leadparagraph{RQ3: How do reasoning effort and model configuration affect performance and cost efficiency?}
We vary Sol's reasoning effort under a common request budget and compare
GPT configurations' API cost efficiency. We compare Qwen sizes and dense versus mixture-of-experts architectures as configuration comparisons, not causal estimates of parameter-count effects.

\subsection{Instances, Agents and Models}
\label{sec:study-subjects}

From the enriched instances, we exclude those limited to documentation adjustments
or surface-level interface changes and manually select 83 reviewed instances
from 11 Python projects as the common cohort for all three tasks and
Agent--model combinations. This is a curated
set, not a random sample of candidate commits; RQ1 examines the broader
construction output separately.

The evaluated systems are Mini-SWE-Agent 2.4.6, Codex 0.147.0 and Claude Code
2.1.233, distinct from the SWE-agent-based construction Agents.
Table~\ref{tab:study-setup} lists the ten models and their reasoning settings;
Sol, Luna and Terra denote the \texttt{gpt-5.6-*} service models used here.
Each of the three Agents is evaluated with every model configuration on the
same 83 instances for all three tasks.
Including Sol's additional effort settings, the twelve configurations yield
$83\times3\times3\times12=8{,}964$ evaluations: 7,470 in the main comparison
and 1,494 in the additional Sol effort comparison.

\begin{table}[htbp]
\caption{Agent--model combinations and shared settings.}
\label{tab:study-setup}
\begingroup
\small
\setlength{\tabcolsep}{4pt}
\renewcommand{\arraystretch}{1.0}
\begin{tabular}{@{}>{\raggedright\arraybackslash}p{0.28\linewidth} >{\raggedright\arraybackslash}p{\dimexpr.72\linewidth-8pt\relax}@{}}
\toprule
\textbf{Setting} & \textbf{Configuration} \\
\midrule
\mbox{Agents (all configurations)} & Mini-SWE-Agent; Codex; Claude Code \\
Sol, Luna, Terra & Medium; Sol additionally low/high \\
Qwen3.8 & 27B: native medium thinking \\
Qwen3.6 & 27B, 35B-A3B: native thinking \\
Qwen3.5 & 9B, 27B, 35B-A3B, 122B-A10B: native thinking \\
Common budgets & 100 main requests; 262,144 context tokens; 8,192 output tokens/request \\
\bottomrule
\end{tabular}
\endgroup

\end{table}

\subsection{Execution and Comparison Protocol}
\label{sec:study-protocol}

For each task and instance, Agents receive identical descriptions and permitted
evidence under Section~\ref{sec:task-construction}'s information boundaries;
the evaluator retains the task-specific private reference material. Agents keep
their native tools and context management, using isolated workspaces without
access to one another's solutions.

Table~\ref{tab:study-setup} specifies the common context, output and main-request
budgets; requests, not shell commands, are counted. Temperature is zero, with
Agent time limits of 2\,h for localization and 4\,h for repair and test generation.
Localization allows up to two additional format-recovery requests of up to 4,096 output tokens each, using the existing record without renewed repository exploration.

We retain one effective attempt per instance--Agent--configuration combination,
not the best retry. Infrastructure interruptions are recovered separately from
capability failures; budget-limited, invalid and unsuccessful runs remain in
the denominator. A verified repair can still succeed after request-budget
exhaustion. Tasks are scored separately, without a composite ranking.

\subsection{Evaluation Metrics}
\label{sec:study-metrics}

For each Agent--model pair, sums cover $N=83$ instances, and $\mathbf{1}[\cdot]$ is an indicator.

\begingroup\predisplaypenalty=0
\leadparagraph{Bug detection and localization.}
Predicted and reference root-location sets $P_i$ and $G_i$ use file--side--line
coordinates, with deleted roots on the parent side. Macro line F1 is primary;
exact-set match and any-hit rates supplement it:
\small
\begin{equation}
\mathrm{MacroF1}=\frac{1}{N}\sum_i\frac{2|P_i\cap G_i|}{|P_i|+|G_i|},\enspace
\mathrm{ExactMatch}=\frac{1}{N}\sum_i\mathbf{1}[P_i=G_i],\enspace
\mathrm{AnyHit}=\frac{1}{N}\sum_i\mathbf{1}[P_i\cap G_i\ne\varnothing].
\label{eq:localization-f1}
\end{equation}
\endgroup
Missing or invalid submissions score zero. Detection recall is
$N^{-1}\sum_i\mathbf{1}[\text{bug reported on }i]$; this positive-only cohort
does not measure false alarms on bug-free commits.

\leadparagraph{Bug repair.}
Private acceptance sets $r_i=1$ if all targeted failures are resolved and
previously passing cases in scope remain passing; otherwise $r_i=0$:
\begin{equation}
\mathrm{RepairCount}=\sum_{i=1}^{N}r_i,\qquad
\mathrm{RepairSuccess}=\frac{1}{N}\sum_{i=1}^{N}r_i.
\end{equation}

\leadparagraph{Regression-test generation.}
For a compliant, completed run, stable regression validation sets $s_i=1$
for matching cases and repeatable
outcomes across two runs per state ($I$, $I+R$), with at least one behavioral
fail-to-pass case and all cases passing on $I+R$; otherwise $s_i=0$.
Import, collection and setup errors or skipped cases do not qualify. Let $D_i$ be matched
fail-to-pass cases in the first execution pair and $C_i$ the generated cases
recorded on $I+R$ in that pair:
\begin{equation}
\mathrm{TestSuccess}=\frac{1}{N}\sum_{i=1}^{N}s_i,\qquad
\mathrm{CasePrecision}=\frac{1}{N}\sum_{i=1}^{N}\frac{|D_i|}{|C_i|}.
\end{equation}
Test success is primary; case precision ignores repeatability and is zero without evidence or if $C_i=\varnothing$.

Let $L_i$ be annotated Python root lines present in $I$ (excluding deletions),
$E_i$ the lines executed by generated tests, and $H_i\subseteq L_i$ the roots
executed by human-written reference tests. Root-line and reference-test-relative
coverage are
\begin{equation}
\mathrm{RootCov}_i=\frac{|E_i\cap L_i|}{|L_i|},\qquad
\mathrm{RefCov}_i=\frac{|E_i\cap H_i|}{|H_i|}.
\end{equation}
Each is macro-averaged where $s_i=1$, coverage is recorded and its denominator
is nonempty; measured and missing counts accompany these conditional metrics.

\leadparagraph{API cost efficiency.}
Let $c_i$ be API cost in USD and $q_i$ be line F1, $r_i$, or $s_i$ for the
respective task:
\begin{equation}
\mathrm{Yield/USD}=\frac{\sum_{i=1}^{N}q_i}{\sum_{i=1}^{N}c_i}.
\end{equation}
Exact matches per USD instead uses $q_i=\mathbf{1}[P_i=G_i]$.
For the five GPT configurations, public short-context token
rates~\cite{openaiSolPricing,openaiLunaPricing,openaiTerraPricing} price all retained
runs, including failures and submission recovery. Reported cache usage is
preserved; unreported cache input counts as uncached, and unreported response
tokens are excluded. Reference API costs exclude full compute costs and are not invoices; dollar rankings exclude self-hosted Qwen. Equal request
limits do not imply equal-dollar budgets.

\section{Results}
\label{sec:results}

\subsection{RQ1: What executable regression data can EReg-Factory construct?}
\label{sec:construction-results}

\leadparagraph{Construction output.}
Mining the 2023--2025 GH Archive data yields 1,694 candidate records across
34 repositories.
We select 11 Python repositories with relatively large candidate pools and
apply EReg-Factory (Section~\ref{sec:construction}) to 522 eligible
candidate instances. Before execution, 145 other candidates are excluded:
126 lack eligible test nodes and 19 lack available test files. Of the 522
instances, 394 complete test execution on both versions and 172 pass
regression relationship verification.

\begin{figure}[htbp]
\centering
\includegraphics[width=0.885\linewidth]{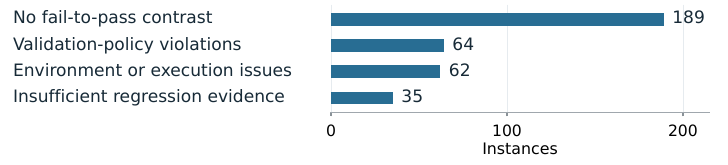}
\caption{Reasons for excluding construction candidates ($n=350$).}
\Description{A horizontal bar chart. Regression construction excludes 189 for absent
fail-to-pass contrast, 64 for validation-policy violations, 62 for environment
or execution issues, and 35 for insufficient regression evidence, including
incomplete downstream records.}
\label{fig:construction-filtering}
\end{figure}

\leadparagraph{What prevents retention?}
The main obstacle is the absence of the required behavioral contrast:
141 candidates do not fail on the introducing version, while 48 fail on the
fixing version. The remaining candidates fall into three categories
(Figure~\ref{fig:construction-filtering}). Validation-policy violations exclude
64 candidates: Agents modify the tested production code, bypass required tests,
or misreport results, contrary to the requirements checked by the Operation
Judge. Environment or execution issues account
for 62 candidates: Agents report unsuccessful environment setup, or execution,
replay or required artifacts remain incomplete. The remaining 35 lack sufficient
regression evidence: results or downstream records do not establish
the proposed introducing--fixing relationship.

Root-cause enrichment retains 153 of the 172 validated instances (89.0\%).
The other 19 instances lack evidence of production-code changes that supports
source-level root-cause annotation. For example, a regression
may arise from a commit's changes to the project's dependency declarations or
versions rather than to the tested production code.

\leadparagraph{Per-instance API cost.}
Table~\ref{tab:construction-costs} reports the mean token usage and API cost
at each automated stage for 153 accepted instances. The model configuration uses DeepSeek
V4 Flash for both Validation Agents and the Root-Cause Agent, and GPT-5.6 Sol
with \texttt{max} effort for the Operation Judges and Regression Verifier.
These automated stages cost \$1.01 per instance on average: \$0.98 for
regression construction and \$0.03 for root-cause analysis.

\begin{table}[htbp]
\caption{Mean token usage and API cost per accepted instance ($n=153$);
k denotes 1,000 tokens.}
\label{tab:construction-costs}
\centering
\small
\setlength{\tabcolsep}{5pt}
\begin{tabular}{@{}lrrrr@{}}
\toprule
\textbf{Stage} & \textbf{Input (k)} & \textbf{Output (k)} & \textbf{Total (k)} & \textbf{USD} \\
\midrule
Introducing Agent & 2,125.2 & 7.1 & 2,132.4 & 0.068 \\
Introducing Judge & 110.3 & 1.4 & 111.6 & 0.416 \\
Fixing Agent & 1,509.3 & 5.1 & 1,514.3 & 0.048 \\
Fixing Judge & 81.0 & 1.1 & 82.2 & 0.314 \\
Regression Verifier & 22.4 & 2.5 & 24.8 & 0.137 \\
\addlinespace[2pt]
Construction Subtotal & 3,848.2 & 17.1 & 3,865.3 & 0.983 \\
\midrule
Root-Cause Agent & 764.9 & 10.6 & 775.6 & 0.027 \\
\midrule
\textbf{Total} & 4,613.1 & 27.8 & 4,640.9 & 1.011 \\
\bottomrule
\end{tabular}

\end{table}

\begin{figure}[htbp]
\centering
\begin{minipage}[t]{0.55\linewidth}
\vspace{0pt}
\includegraphics[width=\linewidth]{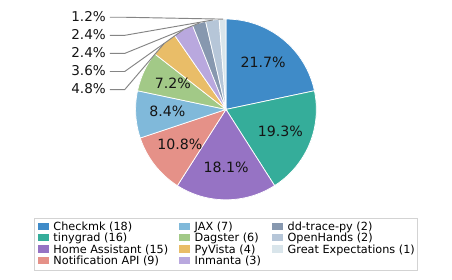}
\caption{Project distribution of the 83 evaluation instances.}
\Description{A solid pie chart of 83 instances from 11 projects, with percentages
inside larger slices and outside smaller slices using leader lines. A compact
three-column legend below lists project names and counts. Segments are ordered clockwise by descending
count: Checkmk 18, tinygrad 16,
Home Assistant 15, Notification API 9, JAX 7, Dagster 6, PyVista 4, Inmanta 3,
dd-trace-py 2, OpenHands 2, and Great Expectations 1.}
\label{fig:construction-project-distribution}
\end{minipage}\hfill
\begin{minipage}[t]{0.42\linewidth}
\vspace{0pt}
\captionsetup{justification=raggedright,singlelinecheck=false}
\captionof{table}{Per-instance root-line and regression-case counts ($n=83$).
Range rows report numbers of instances.}
\label{tab:cohort-characteristics}
\centering
\small
\setlength{\tabcolsep}{4pt}
\begin{tabular}{@{}lrr@{}}
\toprule
\textbf{Count} & \textbf{Root lines} & \textbf{Test cases} \\
\midrule
1 & 26 & 35 \\
2--3 & 29 & 27 \\
4--5 & 9 & 8 \\
6--10 & 6 & 5 \\
$\geq 11$ & 13 & 8 \\
\midrule
Mean & 4.73 & 4.71 \\
\bottomrule
\end{tabular}

\end{minipage}
\end{figure}

\leadparagraph{Evaluation cohort.}
From the enriched instances, we manually select 83 instances across the
11 projects for evaluation (Figure~\ref{fig:construction-project-distribution}).
They contain 393 root-line coordinates and 391 failing regression cases shared
by the three tasks; Table~\ref{tab:cohort-characteristics} summarizes their
per-instance counts.

\begingroup
\definecolor{rqsummaryfill}{HTML}{F2F2F2}
\begin{tcolorbox}[colback=rqsummaryfill,colframe=rqsummaryfill,coltext=black,
  boxrule=0pt,arc=4pt,outer arc=4pt,boxsep=0pt,
  left=12pt,right=12pt,top=5pt,bottom=5pt,
  before skip=5pt,after skip=5pt,fontupper=\normalsize\itshape]
\textbf{Conclusion 1:}
EReg-Factory constructs executable regression instances with independently
reproducible environments for the introducing and fixing versions, verified
introducing--fixing relationships, and regression tests exhibiting fail-to-pass
behavior. Root-cause enrichment adds reviewed source-line locations, repair
patches and causal explanations, providing a shared foundation for bug
localization, repair and regression-test generation.
\end{tcolorbox}
\endgroup

\subsection{RQ2: How do Agents and models differ across the three tasks?}
\label{sec:agent-model-results}

\leadparagraph{Bug repair and regression-test generation show promising results, but precise localization remains challenging.}
Table~\ref{tab:agent-model-results} reports macro-averaged localization F1
(Loc.), repair success rate (Repair), and stable regression-test generation
success rate (Tests) on the same 83 instances for each Agent--model combination.
Among the ten baseline model configurations, Sol with Claude Code repairs
57 instances (68.7\%), and Qwen3.8-27B with Codex generates successful
regression tests for 64 (77.1\%). These outcomes demonstrate successful bug
repair and regression-test generation on a substantial portion of the cohort.
However, no baseline combination exactly recovers the reference root-line
set on more than 17 instances (20.5\%, Sol with Claude Code).

\begin{table}[htbp]
\caption{Agent--model performance (\%; 83 instances per combination).
Column maxima are bold.}
\label{tab:agent-model-results}
\begingroup
\centering\fontsize{7.8}{9.2}\selectfont
\setlength{\tabcolsep}{3.0pt}\renewcommand{\arraystretch}{1.0}
\begin{tabular*}{\linewidth}{@{\extracolsep{\fill}}l*{9}{r}@{}}
\toprule
& \multicolumn{3}{c}{\textbf{Mini-SWE-Agent}} & \multicolumn{3}{c}{\textbf{Codex}} & \multicolumn{3}{c}{\textbf{Claude Code}} \\
\cmidrule(lr){2-4}\cmidrule(lr){5-7}\cmidrule(l){8-10}
\textbf{Model} & \textbf{Loc.} & \textbf{Repair} & \textbf{Tests} & \textbf{Loc.} & \textbf{Repair} & \textbf{Tests} & \textbf{Loc.} & \textbf{Repair} & \textbf{Tests} \\
\midrule
Sol (medium) & 30.9 & 61.4 & 65.1 & \textbf{34.1} & \textbf{67.5} & 67.5 & 31.8 & \textbf{68.7} & 60.2 \\
Luna (medium) & 24.6 & 49.4 & 51.8 & 31.3 & 54.2 & 57.8 & 21.2 & 49.4 & 45.8 \\
Terra (medium) & 24.0 & 60.2 & 60.2 & 26.1 & 59.0 & 59.0 & 27.8 & 61.4 & 56.6 \\
Qwen3.5-122B-A10B & 14.6 & 25.3 & 41.0 & 28.5 & 32.5 & 43.4 & 27.9 & 27.7 & 37.3 \\
Qwen3.5-35B-A3B & 16.7 & 16.9 & 27.7 & 22.0 & 27.7 & 33.7 & 26.3 & 24.1 & 30.1 \\
Qwen3.5-27B & 11.2 & 26.5 & 48.2 & 26.0 & 26.5 & 47.0 & 26.4 & 34.9 & 50.6 \\
Qwen3.5-9B & 6.3 & 13.3 & 27.7 & 24.0 & 19.3 & 22.9 & 24.5 & 16.9 & 24.1 \\
Qwen3.6-35B-A3B & 17.2 & 24.1 & 38.6 & 24.1 & 28.9 & 33.7 & 22.8 & 31.3 & 37.3 \\
Qwen3.6-27B & 23.1 & 36.1 & 51.8 & 23.2 & 41.0 & 49.4 & 24.6 & 36.1 & 53.0 \\
Qwen3.8-27B & 22.0 & 32.5 & 66.3 & 30.0 & 47.0 & \textbf{77.1} & 24.3 & 51.8 & \textbf{75.9} \\
\midrule
\multicolumn{10}{l}{Additional Sol reasoning settings} \\
Sol (low) & \textbf{34.5} & 56.6 & 63.9 & 31.4 & 57.8 & 65.1 & \textbf{35.0} & 62.7 & 61.4 \\
Sol (high) & 29.6 & \textbf{69.9} & \textbf{69.9} & 31.3 & 63.9 & 72.3 & 34.8 & 67.5 & 68.7 \\
\bottomrule
\end{tabular*}
\endgroup

\end{table}

\begin{figure}[htbp]
\centering
\includegraphics[width=\linewidth]{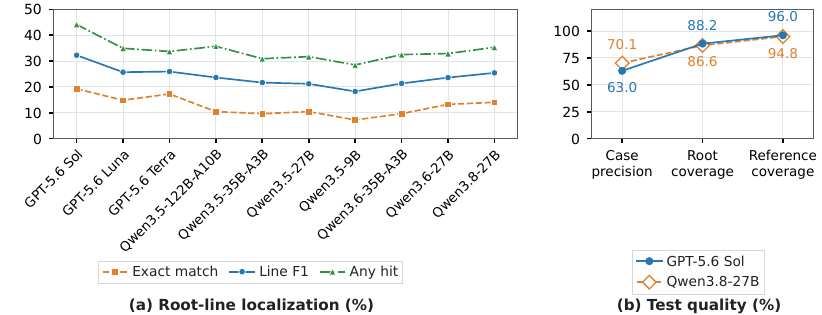}
\caption{Localization and generated-test quality (\%): (a) localization; (b) test precision and root-line coverage.}
\Description{Panel a connects each localization metric across ten models, with full model names on the horizontal axis and percentages on the vertical axis: exact-set match, macro line F1 and any-hit rate
for ten baseline models. Any-hit exceeds exact-set match throughout. Panel b
uses one line per model across three labeled quality metrics to compare GPT-5.6 Sol and Qwen3.8-27B: case precision is 63.0 and 70.1 percent, root coverage
88.2 and 86.6 percent, and reference coverage 96.0 and 94.8 percent. The latter
two measures use recorded coverage from tests that pass regression validation,
not the full cohort.}
\label{fig:task-outcomes}
\end{figure}

The localization measures in Figure~\ref{fig:task-outcomes}a reveal a
substantial gap between finding part of a root cause and identifying its
full extent. Sol's Agent-averaged any-hit rate is 44.2\%, compared with
19.3\% exact-set match and 32.3\% line F1. For Qwen3.8, the corresponding
values are 35.3\%, 14.1\% and 25.4\%. Finding a relevant line is thus
considerably more common than precisely identifying all reference locations.
Any-hit alone would obscure this shortfall; F1 additionally captures both
missed root lines and extraneous reported lines.

\begin{tcolorbox}[colback=black!5,colframe=black!5,boxrule=0pt,arc=4pt,
  outer arc=4pt,boxsep=0pt,left=12pt,right=12pt,top=5pt,bottom=5pt,
  before skip=5pt,after skip=5pt,fontupper=\normalsize\itshape]
\textbf{Conclusion 2:}
Agents can complete substantial portions of repair and regression-test
generation, but precise, complete root-cause localization remains limited.
The gap between any-hit and exact-set match makes this limitation visible.
\end{tcolorbox}

\leadparagraph{Codex leads the Agent averages.}
Table~\ref{tab:aggregate-results} reports equal-weight means across ten baseline
models (a) or three Agents (b); Sol's extra effort settings
enter only RQ3. Codex leads all three Agent averages. Its localization F1
exceeds Claude Code by 1.15 percentage points and Mini-SWE-Agent by 7.87 points.
Repair is nearly tied with Claude Code: 335 versus 334 successes across
830 evaluations (40.36\% versus 40.24\%).

\leadparagraph{Sol leads localization and repair; Qwen3.8 leads test generation.}
Among baseline models, Sol achieves the highest Agent-averaged localization
F1 (32.3\%) and repair success (65.9\%).
Every Agent obtains its highest baseline repair success with Sol.
Qwen3.8 leads test generation at 73.1\%, exceeding Sol by 8.84 percentage
points, and ranks first for each Agent rather than relying on one strong
combination.
Higher task performance need not yield more benefit per dollar;
RQ3 examines this trade-off separately.

\begin{table}[htbp]
\caption{Agent and model averages (\%; 830 and 249 runs per task, respectively).
Bold marks each metric's maximum among Agents or models.}
\label{tab:aggregate-results}
\begingroup
\centering\fontsize{7.2}{8.6}\selectfont
\setlength{\tabcolsep}{1.8pt}\renewcommand{\arraystretch}{1.0}
\begin{tabular*}{\linewidth}{@{\extracolsep{\fill}}lrrr@{\hspace{8pt}}lrrr@{\hspace{8pt}}lrrr@{}}
\toprule
\multicolumn{4}{c}{\textbf{(a) Agent averages}} & \multicolumn{8}{c}{\textbf{(b) Model averages}} \\
\cmidrule(lr){1-4}\cmidrule(l){5-12}
\textbf{Agent} & \textbf{Loc.} & \textbf{Repair} & \textbf{Tests} & \textbf{Model} & \textbf{Loc.} & \textbf{Repair} & \textbf{Tests} & \textbf{Model} & \textbf{Loc.} & \textbf{Repair} & \textbf{Tests} \\
\midrule
Mini-SWE-Agent & 19.1 & 34.6 & 47.8 & Sol (medium) & \textbf{32.3} & \textbf{65.9} & 64.3 & Qwen3.5-27B & 21.2 & 29.3 & 48.6 \\
Codex & \textbf{26.9} & \textbf{40.4} & \textbf{49.2} & Luna (medium) & 25.7 & 51.0 & 51.8 & Qwen3.5-9B & 18.3 & 16.5 & 24.9 \\
Claude Code & 25.8 & 40.2 & 47.1 & Terra (medium) & 26.0 & 60.2 & 58.6 & Qwen3.6-35B-A3B & 21.3 & 28.1 & 36.5 \\
 &  &  &  & Qwen3.5-122B-A10B & 23.6 & 28.5 & 40.6 & Qwen3.6-27B & 23.6 & 37.8 & 51.4 \\
 &  &  &  & Qwen3.5-35B-A3B & 21.7 & 22.9 & 30.5 & Qwen3.8-27B & 25.4 & 43.8 & \textbf{73.1} \\
\bottomrule
\end{tabular*}
\endgroup

\end{table}

\begin{tcolorbox}[colback=black!5,colframe=black!5,boxrule=0pt,arc=4pt,
  outer arc=4pt,boxsep=0pt,left=12pt,right=12pt,top=5pt,bottom=5pt,
  before skip=5pt,after skip=5pt,fontupper=\normalsize\itshape]
\textbf{Conclusion 3:}
Codex achieves the highest average on all three tasks, with repair nearly
tied with Claude Code. Sol leads baseline model localization and repair;
Qwen3.8 leads test generation for all three Agents.
\end{tcolorbox}

\leadparagraph{Generated regression tests cover root-cause code.}
Table~\ref{tab:agent-model-results} reports the main task results for all
combinations. We further compare case precision and root-line coverage in
Figure~\ref{fig:task-outcomes}b for Sol and Qwen3.8, the two baseline models
with the highest test-generation success rates.
Sol and Qwen3.8 achieve macro case precision of 63.0\% and 70.1\%,
respectively. Recorded coverage data from tests that pass regression validation
show mean per-run root-line recall of 88.2\% and 86.6\%. Relative to roots
exercised by the human-written reference tests in each run, mean recall reaches
96.0\% and 94.8\%. Qwen3.8 combines a higher test-generation success rate with
a higher average fraction of generated cases satisfying the regression criterion.
This coverage pattern extends beyond these two models: across the other eight
baseline models, recorded coverage from tests passing regression validation
shows mean root-line recall of 83.8--89.2\% and reference-test-relative recall
of 92.5--96.4\%.

\begin{tcolorbox}[colback=black!5,colframe=black!5,boxrule=0pt,arc=4pt,
  outer arc=4pt,boxsep=0pt,left=12pt,right=12pt,top=5pt,bottom=5pt,
  before skip=5pt,after skip=5pt,fontupper=\normalsize\itshape]
\textbf{Conclusion 4:}
Across models, generated tests that pass regression validation cover, on average,
most root lines and much of the root-cause scope exercised by human-written
reference tests. The recorded coverage shows that these tests not only
distinguish pre- and post-repair behavior but also execute code associated
with the target regression.
\end{tcolorbox}

\subsection{RQ3: How do reasoning effort and model configuration affect performance and cost efficiency?}
\label{sec:configuration-results}

\begin{figure}[htbp]
\centering
\includegraphics[width=0.90\linewidth,trim=0 13bp 0 0,clip]{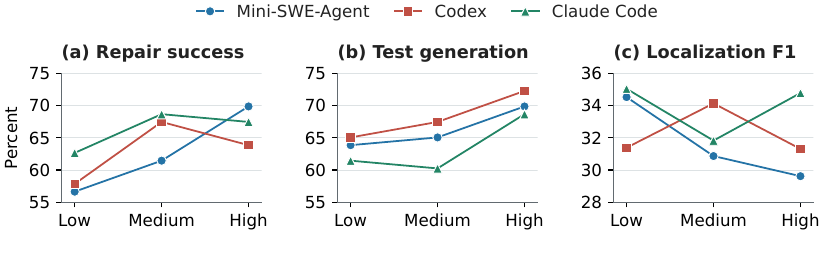}
\caption{Sol reasoning effort and task performance ($n=83$ per point); axis ranges vary by panel.}
\Description{Three line-chart panels compare low, medium and high reasoning
for Mini-SWE-Agent, Codex and Claude Code, ordered as repair, test generation,
and localization. Repair and test-generation axes span 55 to 75 percent;
localization F1 spans 28 to 36 percent. Codex and Claude Code peak at medium
for repair; Mini-SWE-Agent continues to improve. All three Agents reach
their highest test-generation success at high effort. Localization F1
decreases for Mini-SWE-Agent and is non-monotonic for the other two.}
\label{fig:sol-reasoning}
\end{figure}

\leadparagraph{Reasoning effort affects the three tasks differently.}
Figure~\ref{fig:sol-reasoning} compares Sol's three reasoning settings under
the same main-request budget. Average repair success rises from 59.0\% at
low to 65.9\% at medium and 67.1\% at high. Medium captures most of this
gain, with high adding only 1.2 percentage points. Mini-SWE-Agent continues
to improve at high effort, while Codex and Claude Code peak at medium.

Average test-generation success increases from 63.5\% at low to 64.3\% at
medium and 70.3\% at high, a low-to-high gain of 6.8 percentage points.
All three Agents obtain their highest test-generation success at high effort.

Average localization F1 decreases from 33.7\% at low to 32.3\% at medium
and 31.9\% at high. Higher reasoning effort therefore does not improve
recovery of the reference root-line set on average.

\begin{tcolorbox}[colback=black!5,colframe=black!5,boxrule=0pt,arc=4pt,
  outer arc=4pt,boxsep=0pt,left=12pt,right=12pt,top=5pt,bottom=5pt,
  before skip=5pt,after skip=5pt,fontupper=\normalsize\itshape]
\textbf{Conclusion 5:}
Medium reasoning effort captures most of the repair gains; high effort
further improves test generation but not average localization F1.
\end{tcolorbox}

\leadparagraph{Higher success and higher benefit per dollar favor different configurations.}
Table~\ref{tab:evaluation-cost} compares the five GPT settings across
249 runs per task, including unsuccessful attempts. USD/run denotes mean
API cost; Yield/USD divides summed line F1 for localization, or successful-run
counts for repair and test generation, by total cost. Raising Sol from medium
to high multiplies mean repair cost by 2.47, from \$0.374 to \$0.922,
for three additional successes (164 to 167). Test-generation cost rises
1.90-fold, from \$0.242 to \$0.459, for 15 additional successes (160 to 175).
The added effort thus produces a larger observed gain for test generation,
although Sol low delivers the highest benefit per dollar within the Sol
family on all three tasks.

\begin{table}[htbp]
\caption{API cost and task yield across GPT configurations.
Best yields per USD are bold.}
\label{tab:evaluation-cost}
\begingroup
\centering\fontsize{8}{9.2}\selectfont
\setlength{\tabcolsep}{3pt}\renewcommand{\arraystretch}{1.0}
\begin{tabular*}{\linewidth}{@{\extracolsep{\fill}}l*{6}{r}@{}}
\toprule
& \multicolumn{2}{c}{\textbf{Localization}} & \multicolumn{2}{c}{\textbf{Repair}} & \multicolumn{2}{c}{\textbf{Test generation}} \\
\cmidrule(lr){2-3}\cmidrule(lr){4-5}\cmidrule(l){6-7}
\textbf{Configuration} & \textbf{USD/run} & \textbf{Yield/USD} & \textbf{USD/run} & \textbf{Yield/USD} & \textbf{USD/run} & \textbf{Yield/USD} \\
\midrule
Sol low    & 0.1209 & 2.78 & 0.2594 & 2.28 & 0.1850 & 3.43 \\
Sol medium & 0.1873 & 1.72 & 0.3735 & 1.76 & 0.2419 & 2.66 \\
Sol high   & 0.4693 & 0.68 & 0.9216 & 0.73 & 0.4595 & 1.53 \\
Luna medium & 0.0117 & \textbf{21.96} & 0.0232 & \textbf{21.96} & 0.0177 & \textbf{29.19} \\
Terra medium & 0.1245 & 2.08 & 0.2166 & 2.78 & 0.1489 & 3.94 \\
\bottomrule
\end{tabular*}
\endgroup

\end{table}

Luna has the highest yield per dollar in all three tasks, whereas
Sol high achieves the highest GPT repair and test-generation success rates.
Luna's advantage comes from lower token prices, not lower token consumption:
its mean input and output token counts exceed Sol medium's in every task.
The efficiency ranking therefore differs from the capability ranking.
These comparisons describe observed cost--quality trade-offs, not outcomes under a controlled equal-dollar budget.

\begin{tcolorbox}[colback=black!5,colframe=black!5,boxrule=0pt,arc=4pt,
  outer arc=4pt,boxsep=0pt,left=12pt,right=12pt,top=5pt,bottom=5pt,
  before skip=5pt,after skip=5pt,fontupper=\normalsize\itshape]
\textbf{Conclusion 6:}
Higher reasoning effort does not ensure greater benefit per dollar.
Sol high leads GPT repair and test-generation success, while Luna provides
the highest yield on all three tasks.
\end{tcolorbox}

\begin{figure}[htbp]
\centering
\includegraphics[width=0.90\linewidth]{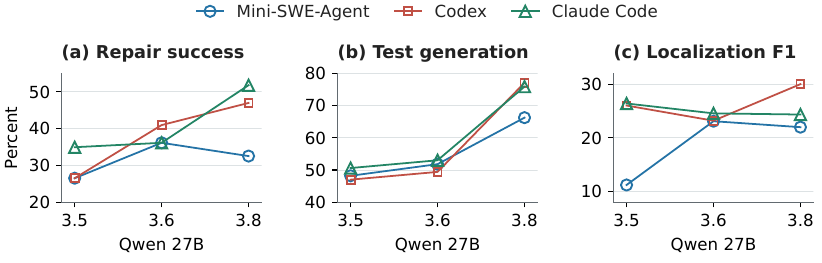}
\caption{Three Qwen 27B generations ($n=83$ per point); axis ranges vary by panel.}
\Description{Three panels compare repair success, test-generation success and
localization F1 across Qwen3.5, Qwen3.6 and Qwen3.8 at 27B. Each panel
contains one curve per Agent. Repair and test-generation success are higher
at Qwen3.8 than Qwen3.5 for every Agent; the largest average gain is in test
generation. Vertical axes span 20 to 55 percent for repair, 40 to 80 percent
for test generation and 8 to 32 percent for localization F1.}
\label{fig:qwen-configurations}
\end{figure}

\leadparagraph{Newer models perform better overall at the same parameter scale.}
To compare model generations at the same parameter scale, we select the
27B configurations of Qwen3.5, Qwen3.6 and Qwen3.8
(Figure~\ref{fig:qwen-configurations}). From Qwen3.5 through 3.6 to 3.8, average repair success
increases from 29.3\% to 37.8\% and 43.8\%; test-generation success rises
from 48.6\% to 51.4\% and 73.1\%. The 3.5-to-3.8 gains are 14.5 and
24.5 percentage points, respectively. Localization F1 also increases,
from 21.2\% to 23.6\% and 25.4\%. The three-Agent curves show that the
repair and test-generation endpoint gains hold for every Agent, not just
the aggregate.

\leadparagraph{Larger total parameter counts do not always yield better performance.}
Within Qwen3.5, the dense 27B model outperforms 9B on all three average
metrics, with repair success of 29.3\% versus 16.5\% and test-generation
success of 48.6\% versus 24.9\%. The larger 122B-A10B model, however, does
not consistently outperform 27B: its localization F1 is higher
(23.6\% versus 21.2\%), but its repair and test-generation success rates
are lower, at 28.5\% and 40.6\%, respectively. Model selection should reflect performance on the target task rather than total
parameter count alone.

\begin{tcolorbox}[colback=black!5,colframe=black!5,boxrule=0pt,arc=4pt,
  outer arc=4pt,boxsep=0pt,left=12pt,right=12pt,top=5pt,bottom=5pt,
  before skip=5pt,after skip=5pt,fontupper=\normalsize\itshape]
\textbf{Conclusion 7:}
At the same parameter scale, newer models perform better overall: average
performance on all three tasks improves across successive generations from
Qwen3.5 through Qwen3.6 to Qwen3.8. A larger total parameter count alone,
however, does not ensure better performance.
\end{tcolorbox}

\FloatBarrier
\section{Discussion and Threats to Validity}
\label{sec:discussion}

\subsection{Discussion}
\label{sec:discussion-implications}

\leadparagraph{Historical regressions as reusable experimental data.}
EReg-Factory turns historical regressions into reusable executable instances
by retaining both versions' replay recipes, validated tests and execution
evidence. Root-cause annotations and tested repairs support complementary
evaluations grounded in the same behavioral difference.

\leadparagraph{Repair and test generation show promise; precise localization needs improvement.}
The strongest combinations show practical promise for repair and test generation
(Section~\ref{sec:agent-model-results}). Identifying relevant code, however,
still falls short of precise, complete root-line recovery, motivating better
links between behavioral reasoning and source changes.

\leadparagraph{Agent--model combinations should be selected by task.}
Relative strengths differ across localization, repair and test generation.
Combinations should therefore be evaluated on the intended task rather than
selected solely from overall Agent or model rankings.

\leadparagraph{Reasoning effort should match the task.}
For Sol, medium effort captures most Agent-averaged repair gains, while high
effort further improves test generation but not mean localization F1
(Section~\ref{sec:configuration-results}). More reasoning is thus not
uniformly beneficial across tasks.

\leadparagraph{Parameter count is not a performance guarantee.}
Larger total parameter counts in Qwen3.5 do not consistently yield better
results, whereas newer Qwen generations improve average results at the same
parameter scale. Total parameter count alone is insufficient for selecting models.

\leadparagraph{Cost efficiency changes the model-selection objective.}
Among GPT configurations, Luna provides the highest yield per dollar, whereas
Sol high leads Agent-averaged repair and test-generation success. Selection
should balance the required performance against API cost.

\subsection{Threats to Validity}
\label{sec:threats}

\leadparagraph{Internal validity.}
Environment reconstruction, automated judgments and annotations can introduce
errors, mitigated through independent replay, evidence checking and two
independent annotators who resolve disagreements through discussion. One
retained effective attempt per combination leaves run-to-run variability
unmeasured.

\leadparagraph{Construct validity.}
Repair acceptance covers targeted failures and previously passing cases in scope,
not complete program correctness. Location matching measures agreement with
reviewed roots; coverage records execution, not comprehensive assertion
adequacy. The positive-only localization cohort does not measure false-alarm
rates on bug-free changes.

\leadparagraph{External validity.}
Our configuration and experiments target Python projects using pytest.
Other languages and testing frameworks may require different environment
setup and test execution mechanisms, and applicability there remains to be
validated.

\section{Related Work}
\label{sec:related-work}

\leadparagraph{Historical defects and executable data.}
SZZ traces fixes to potentially inducing changes~\cite{sliwerski2005szz}.
Defects4J and BugsInPy package reproducible defects for testing and debugging~\cite{just2014defects4j,widyasari2020bugsinpy};
Bugs.jar and ManyBugs broaden the available repair subjects~\cite{saha2018bugsjar,legoues2015manybugs}.
Bears and BugSwarm reproduce failing and passing builds from CI
histories~\cite{madeiral2019bears,tomassi2019bugswarm}.
Building on this emphasis on execution, EReg-Factory follows explicitly linked
introducing--fixing commits and checks fixing-side tests at both historical
states, rather than treating a mined association as a verified regression.

\leadparagraph{Automated environment and task construction.}
Executable repository tasks support evaluation and
training~\cite{jimenez2024swebench,zhang2025swebenchlive,pan2025swegym},
while R2E constructs function-level test harnesses~\cite{jain2024r2e}.
SetUpAgent and Repo2Run automate environment exploration and reusable
setup~\cite{vergopoulos2025setupagent,hu2025repo2run}.
SWE-Factory and SWE-rebench V2 integrate environment preparation with task
validation~\cite{guo2026swefactory,badertdinov2026swerebenchv2};
SWE-smith and R2E-Gym scale task construction through synthetic
generation~\cite{yang2025swesmith,jain2025r2egym}.
Our focus is validating a proposed historical regression: side-specific
environment reconstruction, independent replay, and review of operations and
regression evidence together connect the observed behavior to the introducing
and fixing changes. The retained instances include both replay recipes and
behavioral explanations, supporting reuse beyond repair alone.

\leadparagraph{Fault localization and program repair.}
Spectrum-based, learned, and LLM-guided localizers identify suspicious program
elements~\cite{abreu2006ochiai,li2019deepfl,qin2025soapfl}.
Test-guided repair spans search-based methods and conversational or autonomous
LLM systems~\cite{legoues2012genprog,xia2024chatrepair,bouzenia2025repairagent}.
Repository-level approaches further address code search, planning, and
tool-driven editing~\cite{yang2024sweagent,xia2025agentless,bairi2024codeplan,zhang2024autocoderover}.
EReg-Factory uses repair-supported analysis and human review to annotate
inducing locations separately from repair locations: restoring behavior need
not modify the change that introduced the problem. These annotations support
localization evaluation independently of repair acceptance.

\leadparagraph{Regression tests and behavioral oracles.}
Random and search-based generators construct executable test
suites~\cite{pacheco2007randoop,fraser2011evosuite,lukasczyk2023pynguin}.
LLM approaches generate or improve tests using code context and execution
feedback~\cite{schafer2024testpilot,chen2024chatunitest,alshahwan2024testgenllm};
CoverUp and SymPrompt explicitly guide generation toward coverage~\cite{pizzorno2025coverup,ryan2024symprompt}.
However, coverage alone does not establish fault detection~\cite{shamshiri2015realfaults}:
oracle research examines how assertions distinguish intended from observed
behavior~\cite{barr2015oracleproblem,dinella2022toga,hossain2023neuraloracles}.
SWT-Bench evaluates issue-reproducing tests through fail-to-pass execution and
change coverage~\cite{muendler2024swtbench}.
EReg-Factory supplies validated historical regression tests and reviewed root
locations, linking behavioral discrimination with root-code coverage in the
subsequent test-generation evaluation.

\section{Conclusion}
\label{sec:conclusion}

We presented EReg-Factory, a system for constructing executable regression data
from real-world software history. By combining constrained Agent exploration,
paired environment construction, independent replay, and evidence review, the
system turns candidate introducing--fixing relationships into instances with
validated regression tests and reproducible execution conditions.
Repair-supported analysis and human review enrich these instances with root
locations, repair patches, and causal explanations for localization, repair,
and regression-test generation. Our evaluation shows promising repair and
test-generation performance, while precise root-line localization remains
challenging. Agent--model strengths differ by task; greater reasoning effort or
model size does not uniformly improve performance, and higher success does not
necessarily yield greater benefit per dollar. Together, the construction system
and multi-task evaluation provide a reusable basis for developing executable
data and studying software agents on real historical regressions.

\section*{Data Availability}

An anonymized replication package is available at
\url{https://figshare.com/s/dd865a746d169c689a13}. It includes the EReg-Factory
implementation, constructed datasets and evaluation instances, environment
replay materials, experimental configurations, retained evaluation results,
and analysis scripts.

\clearpage
\bibliographystyle{ACM-Reference-Format}
\bibliography{references}
\end{document}